\documentclass[aip,graphicx]{revtex4-1}
\usepackage{graphicx}
\usepackage{dcolumn}
\usepackage{bm}
\usepackage{amsmath}
\usepackage{amsfonts}
\usepackage{amssymb}
\usepackage{mathrsfs}
\usepackage{color}

\begin{document}


\title{The contribution of magnetic monopoles to  ponderomotive forces in plasmas}

\author{Felipe A. Asenjo}
\email{felipe.asenjo@uai.cl}
\affiliation{Facultad de Ingenier\'ia y Ciencias,
  Universidad Adolfo Ib\'a\~nez, 
  Santiago 7941169, Chile.}

\author{Pablo S. Moya}
\email{pablo.moya@ug.uchile.cl}
\affiliation{Departamento de F\'isica, Facultad de Ciencias,
  Universidad de Chile, Casilla 653, Santiago, Chile.}

\date{\today}

\begin{abstract}
It is well--known that when magnetic monopoles are introduced in
plasma equations the propagation of electromagnetic waves is
modified. This occurs because of Maxwell equations acquire a
symmetrical structure due to the existence of electric and magnetic
charge current densities. In this work we study the nonlinear
phenomena of ponderomotive forces associated to the presence of
magnetic monopoles in a plasma. The ponderomotive force on electric
charges takes into account the symmetrical form of Maxwell equations
in the presence of magnetic charges. It is shown that the general
ponderomotive force on this plasma depends non--trivially on the
magnetic monopoles, through the slowly temporal and spatial variations
of the electromagnetic field amplitudes. The magnetic charges
introduce corrections even if the plasma is unmagnetized. Also, it is
shown that the magnetic monopoles also experience a ponderomotive
force due to the electrons. This force is in the direction of
propagation of the electromagnetic waves.

\end{abstract}

\pacs{14.80.Hv, 52.35.-g, 52.35.Mw}

\keywords{Plasma waves; magnetic monopoles; ponderomotive force.}

\maketitle

\section{Introduction}
The simplest phenomenon of propagation of waves through a plasma can
be studied through the calculation of the dispersion relation of
electromagnetic waves propagating in a medium composed by fixed ions
and moving electrons. Such problem is widely used to introduce
concepts like cutoff and plasma frequencies, as well as, group and
phase velocity of waves. These dispersion properties depend on the
chosen plasma description (kinetic, fluid, magnetohydrodynamics,
etc.), the properties of the media, and on Maxwell
equations. Therefore, if Maxwell equations are modified, then the
propagation of plasma waves is expected to change.

It is well known that Maxwell's equations become symmetric when
electric and magnetic charges (magnetic monopoles) are
theorized~\cite{jackson}. This means that electric and magnetic charge
and current densities appears on every Maxwell equation. One can argue
that the theoretical description of magnetic monopoles is robustly
consistent at many levels in different physical theories
\cite{shnir}. In fact, following quantum mechanical arguments, it is
proved that even the simple existence of one magnetic monopole leads
to the quantization of the elementary electric charge
\cite{dirac1,dirac2}. The minimum coupling strength of the magnetic
monopoles \cite{dirac3} can be calculated to be $\hbar c/(2 e)\simeq
137 e/2$, where $e$ is the value of the elementary electric charge,
$\hbar$ is the reduced Planck constant, and $c$ is the speed of
light. Also, recent models predict that the magnetic monopole mass
should be of the order of $\sim1-10$ GeV \cite{gould}. In the last few
decades, and particularly after the unconfirmed results reported by
Cabrera~\cite{cabrera}, there has been a great interest on the subject
of magnetic monopoles. In addition to different experimental attempts
to detect these particles (see e.g.~\cite{ueno,acharya,patrizii}),
several studies have been published showing (if exist) the role of
magnetic monopoles in different aspects of physics such as
astrophysics~\cite{gould,carrigan}, high energy
physics~\cite{reis,medvedev}, and plasma
physics~\cite{meyer1,meyer2,meyer3,hamilton,moulin}.

If the existence of magnetic monopoles is accepted,
then, from a classical plasma point of view, Maxwell equations
become \cite{jackson,rorh,meyer1}
\begin{eqnarray}
\nabla\cdot{\bf B}&=&4 \pi \mu\, ,\nonumber \\ 
\nabla\cdot {\bf E}&=&-4 \pi e n_e\, ,\nonumber\\
\nabla\times{\bf B}&=&\frac{\partial {\bf E}}{\partial t}-{4\pi}e n_e{\bf v}_e\, ,\nonumber \\
\nabla\times{\bf E}&=&-\frac{\partial {\bf B}}{\partial  t}-{4\pi}\mu {\bf v}_\mu\, ,
\label{eqMaxe}
\end{eqnarray}
where $-e$ and $n_e$ are the electron charge and electron density
respectively, and $\mu$ is the magnetic charge density of the magnetic
monopole fluid. Also, ${\bf v}_e$ is the electron velocity, and ${\bf
  v}_\mu$ is defined as the magnetic monopole fluid velocity (we have
adopted natural units $c=1$). Thus, Maxwell equations can be
interpreted as the electromagnetic field generated by electric and
magnetic fluids, created by their respective constituents. In
addition, from Maxwell equations, we notice that the magnetic monopole
fluid obeys the continuity equation $\partial_t\mu+\nabla\cdot(\mu
{\bf v}_\mu)=0$.

Now we can assume that the above Maxwell equations are coupled to a
plasma composed by electric charges and magnetic monopoles.  We
consider an initial situation where each species is at rest in a
unmagnetized field-free plasma. Let us suppose that there exist other
immobile particles with respective opposite electric and magnetic
charges and with the same densities, which provide the total charge
neutrality of this plasma~\cite{meyer1,meyer2}. Both fluid interact
only through Maxwell equations. Thereby, the electric fluid follow the
momentum equation
 \begin{equation}\label{momelectron}
m_e\left(\frac{\partial}{\partial t}+{\bf v}_e\cdot\nabla\right){\bf v}_e=-e\left({\bf E}+{\bf v}_e\times{\bf B}\right)\, ,
\end{equation}
while the magnetic monopole fluid has its own momentum equation \cite{rorh}
 \begin{equation}\label{mommagneticmonopoles}
m_\mu \left(\frac{\partial}{\partial t}+{\bf v}_\mu \cdot\nabla\right){\bf v}_\mu=\frac{\mu}{n_\mu}\left({\bf B}-{\bf v}_\mu\times{\bf E}\right)\,.
\end{equation}
Here, $m_e$ is the electron mass, and $m_\mu$, $n_\mu$ are the
magnetic monopole mass and magnetic monopole density, respectively.

Due to the above modifications to Maxwell equations and momentum
plasma equations, new effects are expected to emerge in the linear and
nonlinear plasma regimes. In the linear domain, this can be simply
exemplified with the calculation of the dispersion relation for
electromagnetic waves \cite{meyer1}, and how magnetic monopoles
introduce corrections to their propagation. Such analysis has great
theoretical~\cite{moulin,birula} and pedagogical~\cite{meyer1,meyer2}
value. For example, let us consider a monochromatic electromagnetic
wave with electric amplitude $\mathbf{E}_1$ and magnetic amplitude
$\mathbf{B}_1$ propagating through the plasma, disturbing it. At first
order in the perturbations~\cite{birula}, the electron fluid and the
magnetic fluid follow the equations of motion \eqref{momelectron} and
\eqref{mommagneticmonopoles}, $\partial_t {\bf v}_e=-e{\bf E}_1/m_e$
and $\partial_t{\bf v}_\mu=\mu {\bf B}_1/(n_\mu m_\mu)$
respectively. Fourier transforming all the linearized equations with
the form $\exp(i{\bf k}\cdot{\bf r}-i\omega t)$, where $\omega$ is the
frequency and ${\bf k}$ is the wavevector of the electromagnetic wave,
we get ${\bf \tilde v}_e=({-ie}/{m_e\omega}){\bf \tilde E}_1$ and
${\bf \tilde v}_\mu=({i\mu}/{n_\mu m_\mu\omega}){\bf \tilde B}_1$,
respectively, where ${\bf \tilde v}_e$ is the Fourier transform of
${\bf v}_e$, and the same for the other quantities.  Similarly,
Maxwell's equations can be rewritten as ${\bf k}\times{\bf \tilde
  B}_1=\left({\omega_p^2/\omega}-\omega\right){\bf \tilde E}_1$, and
${\bf k}\times{\bf \tilde
  E}_1=\left(\omega-{\omega_m^2}/{\omega}\right){\bf \tilde B}_1$,
where $\omega_p=\sqrt{4\pi e^2 n_{e}/m_e}$ is the electron plasma
frequency, and we introduce the magnetic monopole plasma frequency as
\begin{equation}
\omega_m=\sqrt{\frac{4\pi \mu^2}{n_\mu m_\mu}}\, . 
\label{eqfrequencies}
\end{equation}
Thus, the dispersion relation for a transverse electromagnetic
wave, ${\bf k}\cdot{\bf\tilde E}_1=0$, is given by 
\begin{equation}
\epsilon= \frac{k^2}{\omega^2}=\frac{\left(\omega^2-\omega^2_p\right)\left(\omega^2-\omega^2_m\right)}{\omega^4}\, ,
\label{disprelmono0}
\end{equation}
where $\epsilon$ is the dielectric function. The dispersion
relation~\eqref{disprelmono0} have been obtained in
Refs.~\cite{meyer1,meyer2}. Notice that when magnetic monopoles are
neglected $\omega_m=0$, we recover the usual dispersion relation
$\omega^2=\omega_p^2+ k^2$ for electromagnetic waves in cold plasmas.

\begin{figure}[h!]
 \centering 
\includegraphics[scale=1.25]{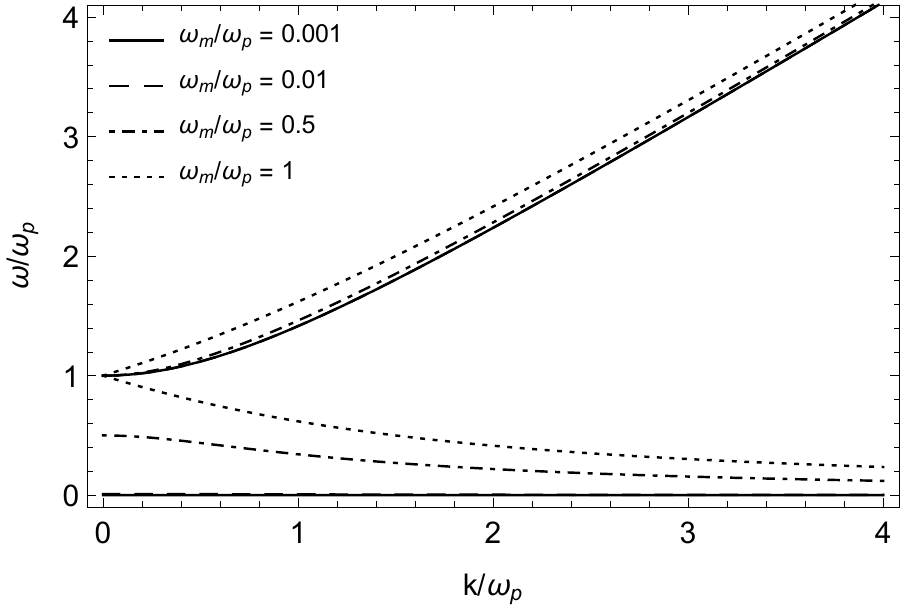}
\caption{Solutions of dispersion relation~\eqref{disprelmono1}. Solid,
  dashed, dot-dashed and dotted lines correspond to $\omega_m/\omega_p
  =$~0.001, 0.01, 0.5, and 0.1, respectively. The dispersion lines
  above $\omega = \omega_p$ correspond to light modes. Those below
  $\omega_p$ are magnetic monopole modes.}
\label{fig1}
\end{figure}

The general solutions, shown in Fig.~\ref{fig1}, for the frequency as
function of $k$ can be straightforwardly obtained from
\eqref{disprelmono0} as
\begin{equation}\label{disprelmono1}
\omega^2_{(\pm)}=\frac{\omega_p^2+\omega_m^2+ k^2\pm\left[\left(\omega_p^2+\omega_m^2+ k^2\right)^2-4\omega_p^2\omega_m^2\right]^{1/2}}{2}\, ,
\end{equation}
where we have used the notation $(\pm)$ to indicate the existence of
two branches. The dispersion relation \eqref{disprelmono0} [or
  \eqref{disprelmono1}] shows that the inclusion of magnetic monopoles
in a plasma formalism introduces non-trivial effects in the
propagation of electromagnetic waves. In the linear regime, this
relation establishes a different form of dispersion of light, that can
be used as an indirect detection method for magnetic charges. From
Fig.~\ref{fig1} we can see how light modes $(+)$ (higher frequency
branches in the figure) are affected by magnetic monopoles
\begin{equation}\label{modomastotal}
\omega^2_{(+)}=\frac{\omega_p^2+\omega_m^2+ k^2+\left[\left(\omega_p^2+\omega_m^2+ k^2\right)^2-4\omega_p^2\omega_m^2\right]^{1/2}}{2}\, .
\end{equation}
Notice that this mode has a cut--off in $\omega_p$, and approaches to
an usual light mode $\omega=k$ as the wavenumber increases. In the
case of small magnetic monopole contribution, $\omega_m\ll k,
\omega_p$, this dispersion relation becomes approximately
\begin{equation}\label{modemasapprox}
\omega^2_{(+)}=\omega_p^2+ k^2+\frac{k^2\omega_m^2}{\omega_p^2+ k^2}\, .
\end{equation}
On the other hand, dispersion relation \eqref{disprelmono1}
establishes the emergence of a new branch $(-)$ due to monopoles. This
branch has the form
\begin{equation}\label{modomenostotal}
\omega^2_{(-)}=\frac{\omega_p^2+\omega_m^2+ k^2-\left[\left(\omega_p^2+\omega_m^2+ k^2\right)^2-4\omega_p^2\omega_m^2\right]^{1/2}}{2}\, ,
\end{equation}
which has a cut--off in $\omega_m$. Also, as the wavenumber increases,
this mode vanishes, as shown in Fig.~\ref{fig1}. For a small magnetic
monopole contribution, $\omega_m\ll k, \omega_p$, this mode becomes
\begin{equation}\label{modemenosapprox}
\omega^2_{(-)}=\frac{\omega_p^2\, \omega_m^2}{\omega_p^2+ k^2}\, .
\end{equation}
This last mode only exist if $\omega_m\neq 0$. Finally, notice from
\eqref{modomastotal} and \eqref{modomenostotal} that
$\omega_+^2>\omega_-^2$, when $\omega_p >\omega_m$.

The above dispersion relations
(Eqs. \eqref{disprelmono0}-\eqref{modemenosapprox}) show how the
linear regime is modified in these kind of plasmas with magnetic
monopoles. We can figure out that the non--linear regime is also
altered. From \eqref{disprelmono0} we see that now
$\partial[\omega^2(\epsilon-1)]/\partial \omega\neq 0$, being opposite
to what occur in a classical electromagnetic wave propagation. This is
a hint indicating that, in the nonlinear regime, new ponderomotive
forces can appear \cite{karp} exclusively due to the presence of
magnetic monopoles. Similar analysis have been performed in
semiconductor plasmas \cite{kisemicond}, quantum plasmas
\cite{jung,nshukla,jung2,lee,poo}, spin quantum plasmas
\cite{brodin1,brodin2}, and in electron-positron plasmas where the
pair annihilation effects are important \cite{ki}.

It is the purpose of this work to obtain and study the contribution of
the magnetic monopoles to the ponderomotive force experienced by the
electrons when a nonlinear electromagnetic wave propagates through a
plasma. The simplest and standard procedure to obtain this nonlinear
force in plasmas is known as the Washimi--Karpman formalism
\cite{karp,karp2,karp3,karp4}, where a large-amplitude electromagnetic
wave induces a ponderomotive force through the spatial and temporal
variations of its amplitude. However, because of the symmetrical form
of Maxwell equations \eqref{eqMaxe} this procedure cannot be directly
applied. Thereby, in order to include the effect of magnetic monopoles
it is necessary to perform, from first principle, a new calculation
for the ponderomotive force. This is performed and detailed in
Sec.~\ref{pond}. To the best of our knowledge, this is the first time
a ponderomotive force is obtained for a symmetrical generalization of
Maxwell equations.  Furthermore, in Sec.~\ref{pondmonop}, a similar
procedure is performed in order to obtain the ponderomotive force
acting on magnetic monopoles mainly owed to the electrons of the
plasma.

\section{Ponderomotive force on electrons owed to the presence of magnetic monopoles}
\label{pond}

Classically, a monochromatic transverse electromagnetic wave
experiences a nonlinear force due to the gradient of the
large--amplitude electric field
\cite{karp,karp2,karp3,karp4,dewar,kentwell,cary}. This force, called
ponderomotive force, can be extended to a more complete form when the
propagation of the electromagnetic wave in a plasma is more general.
When magnetic charges are included this nonlinear force is also
modified. In this section we present the calculation of the
ponderomotive force acting on electrons in a plasma with magnetic
monopoles, where a slowly-varying large--amplitude electromagnetic
wave is propagating in a arbitrary $z$-direction.  Due to the need to
include magnetic monopoles in a symmetrical form in Maxwell equations,
we cannot follow the results derived from the Washimi--Karpman
formalism \cite{karp,karp2,karp3,karp4}. Instead, we derive the
ponderomotive force from first principles. It is important to note
that there exist more complete formalisms \cite{dewar,kentwell,cary}
where general fluid and kinetic effects can be studied into the
framework of a more exhaustive analysis of the ponderomotive
force. However, the lack of symmetry between charged electric and
magnetic particles into Maxwell equations used in those formalisms,
made them not appropriate for the purposes of this work. Nevertheless,
we will show that when magnetic charges are neglected, our results
coincide with those derived from those general formalisms
\cite{kentwell} under the specific imposed conditions of our problem.

Let us consider a transverse slowly large--amplitude electric field
with the form ${\bf E}=\tilde{\bf E}(t,{\bf x}) \exp(ikz - i\omega
t)+{\mbox {c.c.}}$, where c.c. stands for complex conjugate, such that
${\bf E}\cdot\hat z=0$, and the frequency $\omega$ and wavevector
${\bf k} = k\, \hat z$ satisfy the dispersion relation
\eqref{disprelmono1}. The amplitude of the electric field is such that
its gradient is manly in the longitudinal direction $\nabla{\bf
  E}\approx \hat z\,\partial{\bf E}/\partial z$, i.e., the transverse
spatial variations of the electric field are negligible compared with
the longitudinal gradient ($\partial{\bf E}/\partial x, \partial{\bf
  E}/\partial y\ll \partial{\bf E}/\partial z$).  In this way, at
first order in the plasma variables, we find from
Eq.~\eqref{momelectron} that the electron velocity satisfies
\begin{equation}
\left(\frac{\partial}{\partial t}-i\omega\right){\bf v}_{1e}=-\frac{e}{m_e}\tilde{\bf E}\, ,
\end{equation}
showing that the electron velocity is also transverse ${\bf
  v}_{1e}\cdot\hat z=0$. Now, dropping the tilde from the slowly
varying amplitudes for simplicity, and defining the variables
$E_\pm=E_x\pm iE_y$ and $v_{e\pm}=v_{1ex}\pm i v_{1ey}$, we find that
\begin{equation}\label{firstordervelocityelec}
v_{e\pm}=-\frac{e}{m_e\omega}\left(i E_\pm+\frac{1}{\omega}\frac{\partial E_\pm}{\partial t}\right)\, ,
\end{equation} 
where we have used that at the lowest order in
electron velocity amplitude $v_{e\pm}=-ieE_\pm/(m_e\omega)$. A similar
procedure allow us to obtain the first order velocity for magnetic
monopoles. By using \eqref{mommagneticmonopoles}, we get
\begin{equation}\label{firstordervelocitymono}
v_{\mu\pm}=\frac{\mu}{n_\mu m_\mu\omega}\left(i B_\pm+\frac{1}{\omega}\frac{\partial B_\pm}{\partial t}\right)\, ,
\end{equation} 
where $v_{\mu\pm}=v_{1\mu x}\pm i v_{1\mu y}$, defined through the
transverse components of the first order velocity for magnetic
monopoles ${\bf v}_{1\mu}$.

It is now remaining to find the relation between the first order
electric and magnetic fields. Using Maxwell equations \eqref{eqMaxe},
we get
\begin{equation}\label{ecMaxwefirsto}
\nabla\times{\bf E}=-\left(\frac{\partial}{\partial  t}-i\omega\right) {\bf B}-{4\pi}\mu {\bf v}_{1\mu}\, .
\end{equation}
Notice that the last term, proportional to the magnetic monopoles
velocity, is not present in standard Maxwell equations. From
Eq.~\eqref{ecMaxwefirsto}, we obtain that at lowest order 
\begin{equation}\label{exprerelaBE}
B_\pm=\pm \frac{i\, k\,  \omega\,  E_\pm}{\omega^2-\omega_m^2}\, ,
\end{equation}
where we have included the contribution of the magnetic monopole
velocity. Therefore, at first order and with the help of the magnetic
monopole velocity \eqref{firstordervelocitymono}, the magnetic field
solution of \eqref{ecMaxwefirsto} is
\begin{eqnarray}\label{ecMaxwefirsto2}
B_\pm&=&\pm\left(\frac{\omega}{\omega^2-\omega_m^2}\right)\times\nonumber\\
&&\quad\left[ikE_\pm+\frac{\partial E_\pm}{\partial z}+\frac{k}{\omega}\left(\frac{\omega^2+\omega_m^2}{\omega^2-\omega^2_m}\right)\frac{\partial E_\pm}{\partial t} \right]\, .
\end{eqnarray}
When magnetic monopoles are neglected $\omega_m\rightarrow 0$, then
the classical result is recovered as expected.

Now we are in position to calculate the ponderomotive force on
electrons. The nonlinear second order equation of motion for
electrons, obtained from Eq.~\eqref{momelectron}, allow us to get that
ponderomotive force $f_z$ accelerates electrons in the $z$-direction
\cite{brodin1}. Namely,
\begin{equation}
f_z = m_e n_e\left.\left\langle\frac{\partial {\bf v}_{2e}}{\partial
  t}\right\rangle\right|_z= -en_e \left.\left\langle{\bf v}_{1e}\times
{\bf B}\right\rangle\right|_z\, ,
 \end{equation}
where $\langle\, \rangle$ represents the average over the
phase. Note that as the electric and velocity fields
are transverse, there is no contribution from longitudinal quantities
to the ponderomotive force.

From the above expression it is straightforward to show that
\begin{equation}\label{pondforce1}
f_z=-\frac{ie n_e}{2}\left(v_{e+} B_+^*-v_{e+}^* B_+-v_{e-} B_-^*+v_{e-}^* B_-\right)\, .
\end{equation}
Then, to evaluate the ponderomotive force \eqref{pondforce1} we make
use of \eqref{firstordervelocityelec} and
\eqref{ecMaxwefirsto2}. Keeping ourselves in the slowly--varying
regime $E_\pm\,\partial_z E^*_\pm \gg (i/\omega)\,\partial_z
E^*_\pm\,\partial_t E_\pm$, the ponderomotive force on the electron
fluid results to be
\begin{equation}\label{pondforce2}
f_z=-\frac{\omega_p^2}{4\pi\left(\omega^2-\omega_m^2\right)}\left[\frac{\partial}{\partial
    z}+\frac{2k\,
    \omega_m^2}{\omega\left(\omega^2-\omega_m^2\right)}\frac{\partial}{\partial
    t}\right]|{\bf E}|^2\, .
\end{equation}

This is one the main results of this work. Magnetic monopoles
introduce corrections and new sources for the ponderomotive
force. This ponderomotive force, created by a slowly--varying
electromagnetic wave, pushes the electrons locally. This causes charge
separation, thus dragging the whole plasma. The net effect of magnetic
monopoles is to increase the ponderomotive force \eqref{pondforce2},
pushing the electrons faster than in a plasma without them. When
magnetic monopoles are not present, then we recover the standard
ponderomotive force $-({\omega_p^2}/{4\pi\omega^2}){\partial |{\bf
    E}|^2}/{\partial z}$ derived from the Washimi--Karpman formalism
\cite{karp,karp2,karp3,karp4} in a cold electron-ion plasma due to a
transverse electromagnetic wave. In fact, this ponderomotive force
(Eq.~\eqref{pondforce2}) in the vanishing magnetic monopole limit
coincides with the general result derived from Kentwell and Jones
\cite{kentwell} for a fluid plasma theory.

\begin{figure}[h!]
 \centering 
\includegraphics[scale=1.25]{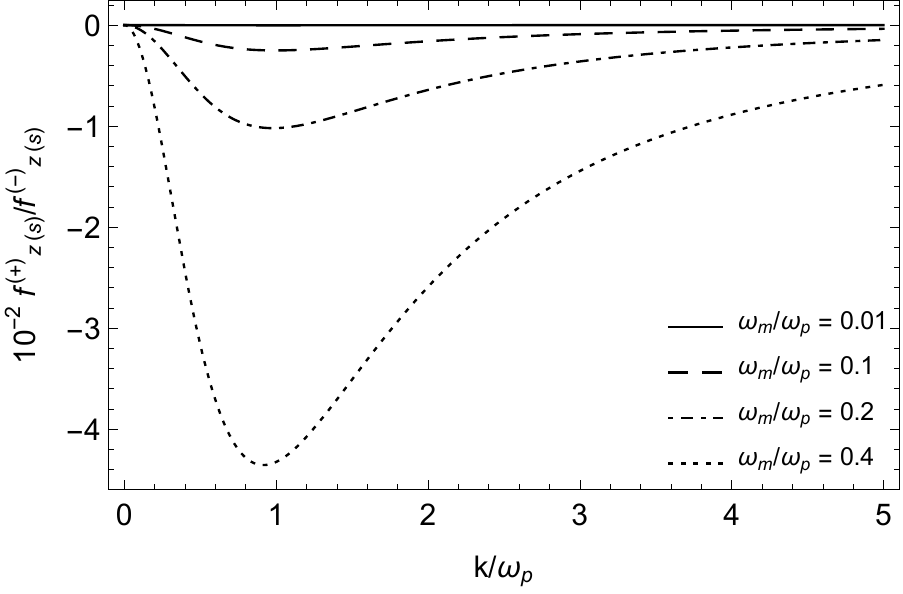}
\caption{Ratio \eqref{comparionponde1} between spatial contributions
  to the ponderomotive forces acting on electrons associated with
  $\omega_{(+)}$ and $\omega_{(-)}$ branches.  Solid, dashed,
  dot-dashed and dotted lines correspond to $\omega_m/\omega_p
  =$~0.01, 0.1, 0.2, and 0.4, respectively.}
\label{fig2}
\end{figure}

Furthermore, due to the two branches predicted by the dispersion relation
\eqref{disprelmono1}, there are two ponderomotive forces
\eqref{pondforce2}, named $f_z^{(+)}$ and $f_z^{(-)}$. Both
ponderomotive forces have the form $f_z={f}_{z(s)}+{f}_{z(t)}$, that
takes into account the spatial $(s)$ and temporal $(t)$ variations of
the amplitude of the electric field. In this way we can estimate the
ratio between the spatial parts of both ponderomotive forces to be
\begin{equation}\label{comparionponde1}
\frac{f_{z(s)}^{(+)}}{f_{z(s)}^{(-)}}= \frac{\omega_{(-)}^2-\omega_m^2}{\omega_{(+)}^2-\omega_m^2}\, .
\end{equation}
As $\omega_{(-)}<\omega_m$, then this ratio is negative (see
Fig.~\ref{fig2}), which implies that the spatial parts of the
ponderomotive forces associated to the two modes accelerate the
electron in opposite directions. Besides, this ratio has a minimum
when $k=\sqrt{\omega_p^2-\omega_m^2}$, which implies that the maximum
value (in absolute terms) of the ratio shown in
Eq.~\eqref{comparionponde1} depends on the relative magnitude between
the magnetic monopole and electron plasma frequencies
$\omega_m/\omega_p$, being larger when $\omega_m$ approaches to
$\omega_p$.  On the other hand, this ratio can be evaluated for small
magnetic monopole contribution $\omega_m\ll k, \omega_p$, resulting to
be
\begin{equation}\label{comparionponde1peque}
\frac{f_{z(s)}^{(+)}}{f_{z(s)}^{(-)}}\approx -\frac{\omega_m^2
  k^2}{\left(\omega_p^2+k^2\right)^2}\rightarrow 0\, .
\end{equation}

\begin{figure}[h!]
 \centering 
\includegraphics[scale=1.25]{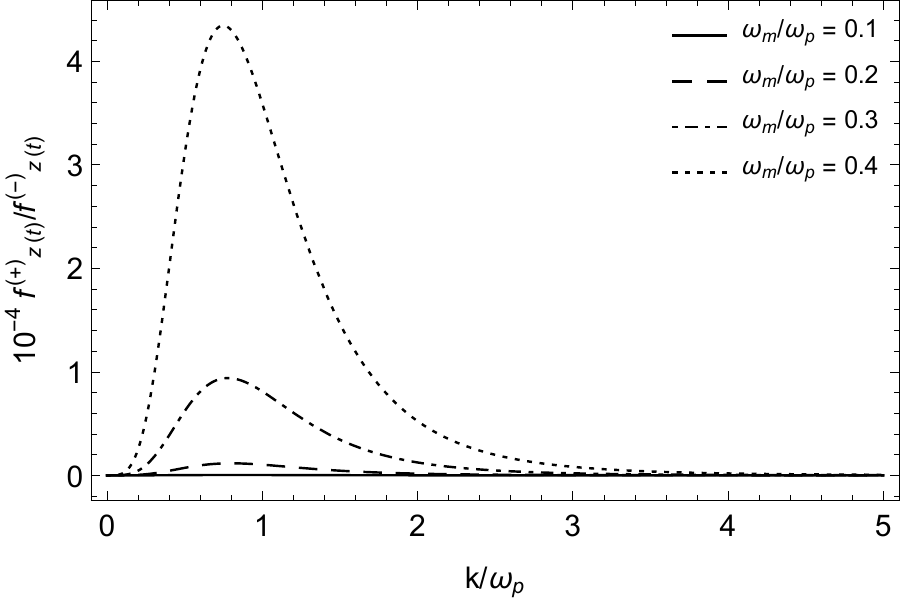}
\caption{Ratio \eqref{comparionponde2} between temporal contributions
  to the ponderomotive forces acting on electrons associated with
  $\omega_{(+)}$ and $\omega_{(-)}$ branches. Solid, dashed,
  dot-dashed and dotted lines correspond to $\omega_m/\omega_p =$~0.1,
  0.2, 0.3, and 0.4, respectively.}
\label{fig3}
\end{figure}

Similarly, the comparison between the temporal parts of
both ponderomotive forces (see Fig.~\ref{fig3}) results to be
\begin{equation}\label{comparionponde2}
\frac{f_{z(t)}^{(+)}}{f_{z(t)}^{(-)}}= \frac{\omega_{(-)}\left(\omega_{(-)}^2-\omega_m^2\right)^2}{\omega_{(+)}\left(\omega_{(+)}^2-\omega_m^2\right)^2}\, ,
\end{equation}
which in the  small magnetic monopole contribution domain becomes
\begin{equation}\label{comparionponde2peque}
\frac{f_{z(t)}^{(+)}}{f_{z(t)}^{(-)}}= \frac{\omega_m^5 k^4 \omega_p}{\left(\omega_p^2+k^2\right)^5}\rightarrow 0\, .
\end{equation}
Ratios \eqref{comparionponde1} and \eqref{comparionponde2} shows that
when $\omega_p>\omega_m$, the ponderomotive forces associated to the
magnetic monopole mode $\omega_-$ are larger in magnitude to those
associated to the light mode $\omega_+$.

To finish the discussion of this section, notice that, in general, the
ponderomotive force \eqref{pondforce2} is different to the result
predicted by the Washimi-Karpman formalism
\cite{karp,karp2,karp3,karp4}. According to it, any classical
high-frequency inhomogeneous electromagnetic wave induces a nonlinear
ponderomotive force given in general by 
\begin{equation}
{\bf f}^{\mbox{\tiny{WK}}}=({\epsilon-1})\,\nabla|{\bf E}|^2/{4\pi}+{{\bf
    k}}\,{\partial_\omega \,[\omega^2 (\epsilon-1)]}\,{\partial_t|{\bf
    E}|^2}/({4\pi\omega^2})\,.
\end{equation}
If we naively would replace the dielectric function
\eqref{disprelmono0} into the Washimi-Karpman formalism for our
modes, it would be obtained that
\begin{equation}
{f}^{\mbox{\tiny{WK}}}_z=-\frac{\omega_p^2}{4\pi\omega^2}\left[\left(1+\frac{\omega_m^2}{\omega_p^2}-\frac{\omega_m^2}{\omega^2}\right)\frac{\partial}{\partial
    z} +\frac{2k\omega_m^2}{\omega^3}\frac{\partial}{\partial
    t}\right]|{\bf E}|^2\, ,
\label{ponderoWK}
\end{equation}
being other components null. The Washimi-Karpman ponderomotive force
${f}^{\mbox{\tiny{WK}}}_z$ is different from the ponderomotive force
\eqref{pondforce2} by several factors. The main reason of the
difference is that the Washimi-Karpman force cannot be applied here as
it does not include the corrections of the magnetic monopole current
density in Maxwell equations. Therefore, the correct ponderomotive
force felt by electrons in the presence of magnetic charges is Eq.
\eqref{pondforce2}, and not the Washimi-Karpman force
\eqref{ponderoWK}.

\section{Ponderomotive force on magnetic monopoles owed to the electrons}
\label{pondmonop}

As the presence of magnetic monopoles induces a ponderomotive force on
electrons, one can argue that electrons produce a similar effect on
magnetic monopoles. An analog analysis to the one in Sec.~\ref{pond}
can be performed to obtain such result. From Amp\`{e}re's law in
Maxwell equations \eqref{eqMaxe} we can obtain at lowest order that
\begin{equation}
E_\pm=\mp\frac{i k \omega B_\pm}{\omega^2-\omega_p^2}\, .
\end{equation}
Notice that this is equivalent to the relation \eqref{exprerelaBE}
making use of the dispersion relation \eqref{disprelmono0}. Thus, at
first order in the variations of the amplitude of the fields, we
obtain that
\begin{eqnarray}\label{ecMaxwefirstoMonop}
E_\pm&=&\mp\left(\frac{\omega}{\omega^2-\omega_p^2}\right)\times\nonumber\\
&&\quad\left[ikB_\pm+\frac{\partial B_\pm}{\partial z}+\frac{k}{\omega}\left(\frac{\omega^2+\omega_p^2}{\omega^2-\omega^2_p}\right)\frac{\partial B_\pm}{\partial t} \right]\, .
\end{eqnarray}
showing the inherent symmetry between electric and magnetic charges,
when it is compared with Eq.~\eqref{ecMaxwefirsto2}.

Following the same procedure outlined in Sec.~\ref{pond}, we obtain
the second--order ponderomotive force, $F_z$, induced on magnetic
monopoles as
\begin{eqnarray}
F_z&=&-\mu \left.\left\langle{\bf  v}_{1\mu}\times {\bf E}\right\rangle\right|_z\nonumber\\
&=&-\frac{i\mu}{2}\left(v_{\mu+} E_+^*-v_{\mu+}^* E_+-v_{\mu -} E_-^*+v_{\mu -}^* E_-\right)\, .
\end{eqnarray}
By using the slowly--varying magnetic monopole velocity deduced in
Eq.~\eqref{firstordervelocitymono}, we finally find that the
longitudinal ponderomotive force on magnetic monopoles, in terms of
the slowly--varying amplitude of the magnetic field, results to be
\begin{equation}\label{podforcemagnmonop}
F_z=-\frac{\omega_m^2}{4\pi\left(\omega^2-\omega_p^2\right)}\left[\frac{\partial}{\partial
    z}+\frac{2k\omega_p^2}{\omega(\omega^2-\omega_p^2)}\frac{\partial}{\partial
    t}\right]|{\bf B}|^2\,.
\end{equation}
The ponderomotive force \eqref{podforcemagnmonop} acts on the
magnetic monopoles, accelerating them along the longitudinal
direction. Notice the symmetry with respect to the ponderomotive force
on electrons \eqref{pondforce2}, where only is needed to perform the
changes $|{\bf E}|^2\longleftrightarrow |{\bf B}|^2$ and $\omega_p
\longleftrightarrow\omega_m$. This is a reflection of the symmetric
form of Maxwell equations.

Anew, we can compare the spatial and temporal parts of the
ponderomotive force \eqref{podforcemagnmonop} (that take into account
the temporal and spatial variations of the magnetic field) for the
light and monopole modes. The ratio of the spatial parts of the
ponderomotive force on magnetic monopoles gives
\begin{equation}
\frac{F_{z(s)}^{(+)}}{F_{z(s)}^{(-)}}=
\frac{\omega_{(-)}^2-\omega_p^2}{\omega_{(+)}^2-\omega_p^2}\approx
-\frac{\omega_p^2}{k^2}\, ,
\end{equation}
and therefore, the spatial parts for the two modes pushes the
monopoles in opposite directions (as $\omega_{(-)}<\omega_p$), in an
analogue fashion to the electron case. Similarly, the ratio of the
temporal parts of the ponderomotive force \eqref{podforcemagnmonop} is
\begin{equation}
\frac{F_{z(t)}^{(+)}}{F_{z(t)}^{(-)}}=
\frac{\omega_{(-)}\left(\omega_{(-)}^2-\omega_p^2\right)^2}{\omega_{(+)}\left(\omega_{(+)}^2-\omega_p^2\right)^2}\approx
\frac{\omega_m\omega_p^5}{k^4(k^2+\omega_p^2)}\, ,
\end{equation}
aiming the temporal parts of the force moves monopoles in the same
direction for the modes. Also, as the wavenumber increases, the
temporal part of the magnetic monopole mode $F_{z(t)}^{(-)}$ becomes
dominant.




\section{Discussion and conclusions}

We have studied how magnetic monopoles modify the linear classical
propagation characteristics of electromagnetic waves in plasmas in the
linear and nonlinear cases, leading to non-trivial effects. When the
nonlinear regime is explored, magnetic charges contribute to the
ponderomotive forces experienced by the electrons. We show that the
calculation of the ponderomotive force cannot follow the standard
Washimi--Karpman procedure, as now it is mandatory to consider the
current density created by the motion of magnetic monopoles.

We found that when considering magnetic charges in the plasma, the
ponderomotive force \eqref{pondforce2} that an electromagnetic waves
exerts on electrons is larger than its standard classical
counterpart. This implies that the electrons in a plasma can move
faster due to the presence of magnetic charges. In fact, if some of
our assumptions are relaxed, for example the constraints to the
spatial variations of the electric field amplitude, then it can be
glanced that the term $\omega^2-\omega_p^2$ in the denominator of
Eq.~\eqref{pondforce2} contributes to the transverse part of the
ponderomotive force, being that part also larger than its classical
counterpart. Therefore, it is possible to argue that magnetic
monopoles can contribute into the self-focusing of a laser in an
electron plasma. This idea will be investigated in future works.
Lastly, we showed that magnetic monopoles can also experience the
ponderomotive force \eqref{podforcemagnmonop} in a plasma. This force
also accelerate the monopoles in a similar fashion to what occurs with
electrons.

Our theoretical results show that, if magnetic charges exist, there is
a variety of effects that may be present when large amplitude
electromagnetic waves propagate through plasmas with no background
magnetic field. As all these effects are impossible to observe without
the presence of magnetic charges, our results suggest a possible
procedure to detect these, until today, elusive particles.

\begin{acknowledgments}
P.S.M. is grateful for the support of CONICyT Chile through FONDECyT
grant No. 11150055, and Conicyt PIA project ACT1405.
\end{acknowledgments}

\end{document}